\documentstyle[11pt,aaspp4]{article}
\slugcomment{To appear in The Astrophysical Journal.}

\lefthead{Yun \& Scoville}
\righthead{CO (J = 4 -- 3) Observations of FSC~15307+3252}

\newbox\grsign \setbox\grsign=\hbox{$>$} \newdimen\grdimen 
\grdimen=\ht\grsign
\newbox\laxbox \newbox\gaxbox
\setbox\gaxbox=\hbox{\raise.5ex\hbox{$>$}\llap
     {\lower.5ex\hbox{$\sim$}}}\ht1=\grdimen\dp1=0pt
\setbox\laxbox=\hbox{\raise.5ex\hbox{$<$}\llap
     {\lower.5ex\hbox{$\sim$}}}\ht2=\grdimen\dp2=0pt

\def\kms    {\ifmmode{{\rm ~km~s}^{-1}}\else{~km~s$^{-1}$}\fi}

\def\deg      {{\ifmmode^\circ\else$^\circ$\fi} } %%% Overwrites TeX \deg

\newbox\grsign \setbox\grsign=\hbox{$>$} \newdimen\grdimen 
\grdimen=\ht\grsign
\newbox\simlessbox \newbox\simgreatbox
\setbox\simgreatbox=\hbox{\raise.5ex\hbox{$>$}\llap
     {\lower.5ex\hbox{$\sim$}}}\ht1=\grdimen\dp1=0pt
\setbox\simlessbox=\hbox{\raise.5ex\hbox{$<$}\llap
     {\lower.5ex\hbox{$\sim$}}}\ht2=\grdimen\dp2=0pt
\def\simgreat{\mathrel{\copy\simgreatbox}}
\def\simless{\mathrel{\copy\simlessbox}}

\begin{document}

\title{CO (J = 4 $\rightarrow$ 3) and 650 $\mu$m Continuum Observations \\
of the z=0.93 Hyperluminous Infrared Galaxy FSC~15307+3252}

\author{M. S. Yun}
\affil{National Radio Astronomy Observatory,\footnote{The National 
Radio Astronomy Observatory is 
a facility of the National Science Foundation operated under 
cooperative agreement by Associated Universities, Inc.} 
	P.O. Box 0, Socorro, NM 87801}
\and 

\author{N. Z. Scoville}
\affil{California Institute of Technology, Pasadena, CA 91125}
 
\begin{abstract}

We report the results of our CO $J=4\rightarrow 3$ line and rest frame
650 $\mu$m continuum observations of the z=0.93 hyperluminous
infrared galaxy FSC~15307+3252 using the Owens Valley Millimeter
Array.  No line or continuum emission was detected, but the
derived limits
provide a useful constraint on the temperature, emissivity, and mass
of the cold dust associated with FSC~15307+3252 and its
molecular gas content.  

The 3$\sigma$ upper limit on the velocity integrated
CO (4--3) line flux is 1.6 Jy \kms\ (for $\Delta V=300$ \kms).
This corresponds to a surprisingly small total molecular gas
mass limit of $5\times 10^9 h^{-2}~M_\odot$ 
for this galaxy with infrared luminosity $L_{FIR} > 10^{13} L_\odot$.  
Combined with existing photometry data, our 3$\sigma$ upper limit
of 5.1 mJy for the 239 GHz (650 $\mu$m rest wavelength) continuum 
flux yields a total dust mass of 0.4-1.5 times $10^8~M_\odot$.
The CO luminosity (thus molecular gas content)
and the resulting gas-to-dust ratio are lower than 
the values typical for the more gas-rich infrared
galaxies, but they are within the observed ranges.  
On the other hand, FSC~15307+3252 has a
dust content and infrared luminosity 40 and 200 times larger
than the infrared bright
elliptical-like galaxies NGC~1275 and Cygnus~A.

The FIR luminosity to dust mass ratios, $L_{FIR}/M_{dust}$,
for all three galaxies hosting a powerful AGN (FSC~10214+4724, 
FSC~15307+3252, Cygnus~A) are larger than reasonably expected 
for a galaxy dominated by a starburst and four times 
larger than Arp~220.  Therefore
the bulk of the observe FIR luminosity in these galaxies
is likely powered by their luminous active nucleus.

\end{abstract}


\section{Introduction}

The early evolution of galaxies may be extremely rapid with a 
burst of high luminosity.  Optical searches for such ultraluminous 
proto-galaxies have been unsuccessful, indicating that
such activity is obscured by dust extinction or by Lyman-continuum 
absorption (internal or intervening) and therefore 
not optically visible (e.g. \cite{djor93}).  Alternatively,
this may also mean that most galaxies
do not have an early high luminosity phase.  Millimeter wavelength
CO and submillimeter dust continuum searches offer an
alternative, extinction-free way of detecting high redshift
proto-galaxies (see Yun \& Scoville 1996, Evans et al. 1996, \cite{smail97,hacking97}). 

The first detection of CO emission from a cosmological distance
was reported by Brown \& Vanden Bout (1991) and by Solomon et al. 
(1992a) in the most distant infrared galaxy detected by the IRAS
survey, FSC~10214+4724 at z=2.3 (\cite{rowan91}).
The extended nature of the CO emission was later revealed by 
high resolution imaging (\cite{sco95,downes97}).   
In a unique study exploiting the magnifying 
property of gravitational lensing, the total gas mass and the detailed
distribution of molecular gas in the z=2.6 quasar H~1413+117 were
determined (Yun et al. 1997), and the derived gas 
properties are indeed remarkably similar to the IR luminous galaxies 
in the local universe ($d_{CO}\simless$ 1 kpc, $n\ge 10^4$ cm$^{-3}$,
$T_{ex}\ge$ 50-100 K).  The compact and concentrated gas 
distribution suggests that the rapid metal enrichment of  
interstellar medium by a massive starburst may explain 
the strong CO emission.  Redshifted CO emission has
subsequently been detected in the z=4.7 quasar BR~1202$-$0725 
(\cite{ohta96,omont96}) and in the $z=2.4$ radio galaxy 53W002
(\cite{sco97a}).

The hyperluminous infrared galaxy FSC~15307+3252 at 
$z_{opt}=0.9257\pm 0.0012$ is the
second highest redshift object detected by IRAS and one of the most
luminous galaxies known ($\nu L_{60\mu}=4\times 10^{12} L_\odot$, 
Cutri et al. 1994).  Both the optical spectrum (Seyfert 2)
and the radio continuum power
(log $P_{1.4GHz}=25.1$ W Hz$^{-1}$ -- from the VLA FIRST Survey 
by \cite{becker95}) suggest some nuclear
activity, and the presence of polarized, broad optical emission lines
suggests that this galaxy may host an obscured
QSO (\cite{hines95,liu96}).
The observed properties of FSC~15307+3252, including the 
spectral energy distribution dominated by the infrared, 
are nearly identical to the more distant IRAS galaxy 
FSC~10214+4724, where bright (lensed) CO emission has already 
been detected.  In general, strong dust emission is 
one of the best indicators of a large molecular gas content, and
the large 60 $\mu$m luminosity and infrared-to-optical luminosity
ratio suggest that FSC~15307+3252 is also a high redshift 
dusty, gas-rich galaxy.  Therefore we have conducted 
a search for redshifted CO (4--3) emission and
submillimeter continuum emission at $\lambda_{rest}=650~\mu$m using the 
Owens Valley Millimeter Array between December 1996 and
February 1997.  Here we report the results of our search and
the inferences on the dust and molecular gas content in this 
hyperluminous infrared galaxy.
\bigskip

\section{Observations}

FSC~15307+3252 was observed using the Owens Valley Millimeter
Array in the low and equatorial configurations between December 
1996 and February 1997. The array consists of six 10.4 m
telescopes, and the longest baseline observed was 120 m.
The redshifted CO (4--3) transition occurs at 239.41 GHz, 
and the primary beam of the 10.4-m telescope is 24.9\arcsec\ 
(140 kpc at $D_A=1.16$ Gpc\footnote{We adopt $H_\circ=75$ 
Mpc~[km s$^{-1}$]$^{-1}$ and $q_\circ=0.5$ throughout this paper.})
at this frequency.  
The system temperatures were typically 600--900 K in the signal
side-band corrected for antenna and atmospheric losses and 
for atmospheric incoherence.  Spectral resolution was 
provided by a digital correlator
configured with 120$\times$4 MHz channels (5.0 \kms), 
covering a total velocity range of 600 km s$^{-1}$.  
The 239.4 GHz continuum emission was measured using a 1 GHz 
bandwidth analog correlator, and the 3mm continuum emission
centered on 102.1 GHz was measured simultaneously using
the 3mm SIS receivers ($T_{sys}$ = 200--400 K).  The total
usable integration time on source was 19 hours distributed over 4
tracks in the two different telescope configurations.  
The nearby quasar 1611+343 (1.0 \& 0.8 Jy at 102 \& 239 GHz) was used 
to track the phase and gain variations. Neptune (T$_b$=90 K) 
and 3C~273 were observed for absolute flux calibration -- see 
Table 1 for the observation summary. 
The data were calibrated using the standard Owens Valley array
program mma (\cite{sco92}) and mapped using DIFMAP
(\cite{she94}) and the NRAO AIPS package. 
\bigskip

\section{Results}

\subsection{CO (4--3) Emission}

We searched for CO $J = 4 \rightarrow 3$ ($\nu_{rest}=461.040811$ GHz)
line emission in the redshift range
z=0.9240-0.9275, but no CO emission was detected.  
The rms noise in each 80 MHz ($\Delta V=100$ \kms) channel was 
9 mJy beam$^{-1}$ (see Figure 1).  The derived upper limit on
the molecular gas mass is $M_{H_2}(3\sigma)\le 5.0\times 
10^9 h^{-2} M_\odot$, assuming a CO line width of 300 \kms\
and $M_{H_2}/L'_{CO} = 4~M_\odot$ [K km s$^{-1}$ pc$^2$]$^{-1}$.
This is about 1/2 or 1/3 of the amount of molecular gas 
found in the ultraluminous infrared galaxies 
such as Arp~220 (see \cite{san91,sco97b}) and a factor two 
better than the upper 
limits reported by Evans et al. (1998) from their 
IRAM 30-m telescope observations.  It is possible that  
the CO emission occurs at a redshift significantly different
from the UV/optical emission line redshift we searched.  
However, the near-infrared
spectroscopy of the narrow emission lines by Evans et al. (1998) 
confirms the redshift determined earlier by Cutri et al. (1994),
and it is therefore likely that the CO emission in FSC~15307+3252 is 
intrinsically weak.  

\subsection{Continuum Emission}

The continuum emission at 102 GHz and 239 GHz were measured
simultaneously with the
spectral line data using the analog continuum correlator
with 1 GHz bandwidth.  The rms noise estimated from the 
natural-weighted image was 0.43 mJy beam$^{-1}$ at
102 GHz and 1.7 mJy beam$^{-1}$ at 239 GHz.
The spectral energy distribution (SED) of FSC~15307+3252 is
shown in Figure~2 including the 3$\sigma$ upper limits of
1.3 mJy and 5.1 mJy at $\lambda_{rest}$
of 1500 $\mu$m and 650 $\mu$m and
the photometric data from Cutri et al. (1994) and
Becker et al. (1995).  

The spectral index between the
IRAS measurement and our 650 $\mu$m upper limit,
$\alpha$(52--650$\mu$m) $>$1.8, is not large enough
to rule out the possibility of a spectral break
due to self-absorbed synchrotron emission. 
If the spectral break occurs at $\lambda\simgreat 
100~\mu$m, however, it would clearly exceed the theoretical limit 
($\alpha=2.5$) even if the 650 $\mu$m flux is close to the
3$\sigma$ upper limit reported here, leaving thermal dust
emission from a large amount of dust as the only reasonable
explanation for the large FIR/submillimeter luminosity.  The 
observed FIR and submillimeter SED can be described well by 
the thermal continuum emission from $T_{dust}= 50$ K dust 
with emissivity $\beta=1.5$ (see below for a further discussion).
\bigskip

\section {Discussion} 

\subsection {Dust Temperature, Emissivity, and Mass}

The flux density, $S_\nu$, from a cloud at a distance
$D$ containing $N$ spherical dust grains each of cross-section
$\sigma$, temperature $T$, and emissivity $Q(\nu)$, is given by 
\begin{equation}
S_\nu = N[\sigma/D^2] Q(\nu) B(\nu,T)
\end{equation}
The dust cloud mass can be obtained as the product of $N$,
the assumed mean dust mass density $\rho$, and the mean volume 
of each dust particle $v$.  The dust volume $v$ 
can be related to the grain cross-section $\sigma$ in Eq. 1,
and the dust mass can be obtained from the observed flux as
\begin{equation}
M_{dust}=[S_\nu D^2/B(\nu,T)][(4/3)a/Q(\nu)]\rho
\end{equation}
where $a$ is the grain radius (see Hildebrand 1983).  The 
mass opacity coefficient $\kappa_\nu \equiv [(4/3)a\rho/Q(\nu)]^{-1}$
may be strongly dependent on frequency and dust composition,
and its estimate and resulting dust mass are uncertain 
by a factor of at least three (see \cite{hughes93}).  Adopting 
$\kappa_\nu = 0.1~{\rm g}^{-1}~{\rm cm}^2$ at $\lambda=250\mu$m 
(Hildebrand 1983), Eq. 2 becomes
\begin{equation}
M_{dust}=180 {{S_{250\mu m}D_L^2}\over{(1+z)}}[exp(57.6/T_{dust}) -1]~M_\odot
\end{equation}
where $S_{250\mu m}$ is the 250 $\mu$m flux density in Jy and $D_L$ is the luminosity distance in Mpc.  

The continuum flux limits we obtained at millimeter and 
submillimeter wavelengths provide a useful constraint on 
the temperature, emissivity, and mass
of the cold dust associated with FSC~15307+3252.  
%The submillimeter
%spectrum of cool dust clouds can be described as the product of
%the black-body intensity, $B(\nu,T)$, and an emissivity $Q(\nu)$
%which varies as $\nu^\beta$, where $\beta$ increases from near
%unity at $\lambda\simless200~\mu$m to $\ge2$ at 
%$\lambda\simless1000~\mu$m (see Hildebrand 1983).  
For the gas rich infrared luminous galaxies
such as Arp~220, the observed SED is consistent with the 
optical depth reaching unity near $\lambda\sim100~\mu$m
(\cite{sco91,sol97}), and a high opacity at these wavelengths
is also suggested by the faintness of 158 $\mu$m [C II]
fine structure line emission observed in such galaxies 
(\cite{malhotra97,fischer98}).  We therefore model the observed
FIR and submillimeter SED as the combination of a pure black-body
spectrum at short wavelengths and a grey-body spectrum with
emissivity $\beta$ between 1 and 2 for the long wavelengths,
with the transition occurring between 100 and 200 $\mu$m.

Three different ``good" dust emission models for FSC~15307+3252 
are listed in Table~2 and plotted in Figure 2.  The IRAS 60 and 
100 $\mu$m ($\lambda_{rest}$ = 31 and 52 $\mu$m) data 
are well fit by a pure black-body spectrum ($\beta=0$) with
dust temperature of 85 K.  At longer wavelengths, 
dust emission from 35 K dust with emissivity $\beta = 2.0$ 
(Model A) is consistent with the $\lambda_{rest}=650~\mu$m 
upper limit and the 85 K black-body spectrum at shorter wavelengths.
This also represents the maximum dust mass case ($M_{dust}=
1.5\times 10^8 M_\odot$) among the
variety of models considered because it has the lowest $T_{dust}$
and predicts the largest 250 $\mu$m flux (see Eq.~3).
Other $\beta = 2.0$ models with $T_{dust}\le30$ K and 
bound by the $\lambda_{rest}=650~\mu$m upper limit do not
connect smoothly with the 85 K black-body spectrum (IRAS data).  

The actual observed spectra of luminous infrared 
galaxies such as Arp~220 and FSC~10214+4724 are well described 
by a dust model with emissivity $\beta \sim 1.5$ (see Table~3).
A $\beta = 1.5$ model and $T_{dust}=50$ K (Model B) is also 
consistent with the $\lambda_{rest}=650~\mu$m upper limit and the
85 K black-body spectrum at shorter wavelengths.  The 
derived dust mass is about 40\% smaller than Model~A
mainly due to higher dust temperature.  

Given the limited data available, an exhaustive search for the best
model is not warranted, and the ``good" dust models shown here are
not unique.  Nevertheless some meaningful limits on the dust
temperature and emissivity can be set by the observations.  
For example, a dust spectrum with $T_{dust}=50$ K and 
emissivity $\beta = 1.0$ produces too much 650 $\mu$m emission
when connected smoothly with the black-body spectrum at
shorter wavelengths.  Similarly, model spectra
with $T_{dust}=40$ K and $\beta\le1.5$ are excluded by the data.
Using these constraints, a minimum dust mass model with the
highest $T_{dust}$ allowed by the IRAS data points and the
largest $\beta$ physically meaningful may be constructed and
is shown as Model C in Table~2 and Figure~2.  The minimum dust mass 
for FSC~15307+3252 (Model C) is $3.5 \times 10^7~M_\odot$,
which is about 1/4 of the maximum dust mass in Model~A.

In summary, a wide range of possible dust models considered
here have produced a relatively narrow range of total dust
mass for FSC~15307+3252.  A few clarifying remarks
should made here, however.  First of all, 
our modeling offers a robust estimate for the {\it total} dust 
mass as the cold dust component in the submillimeter band 
dominates the total dust mass.  
This is clearly demonstrated by the fact that the apparent 
difference in the observed flux density between the 85 K 
and 200 K dust component is only a factor of three near
$\nu_{rest}= 10^{12.8}$ Hz (see Figure~2)
but the warmer component accounts for only about 1\% in dust mass
because of strong temperature dependence on flux density.  Secondly,
while it is conceivable that a much larger quantity of extremely cold dust
may be present in this galaxy for the same reason, the dust models 
listed in Table~2 are nevertheless the most relevant in understanding
the nature of FSC~15307+3252.  In theory, there could be an
arbitrarily large amount of dust at or near the cosmic 
background temperature ($T_{CMB}=5.3$~K) because such a dust
component would not contribute much to the observed SED.  
On the other hand, such cold dust, if present, would not
be found in the area of the galaxy associated with the 
strong heating, and it would be irrelevant to the 
physics of energetic processes observed in this galaxy.  
Therefore the relevant maximum total dust mass for 
FSC~15307+3252 is about $1.5\times 10^8 M_\odot$.

\subsection {Comparison with Ultraluminous Infrared Galaxies}

The extremely dusty nature of FSC~15307+3252 is easily seen 
when it is compared to the two well studied infrared luminous
galaxies FSC~10214+4724 and Arp~220.  As summarized in
Table~3, the FIR luminosity of FSC~15307+3252 is 10 times larger
than Arp~220 while their optical B band luminosities are comparable.  The
$L_{FIR}/L_B$ ratio of 650 is about four times larger than both
FSC~10214+4724 and Arp~220, and the dominance of the FIR in its 
bolometric luminosity is the most extreme among all known objects.  

For optically thick black-body emission (at $\lambda \simless 100~\mu$m) 
originating from a single source with a radius $R$, the source 
size can be determined as $R=[{{S_\nu D_L^2}\over{\pi B(\nu,T)}}]^{1/2}$  
-- for a distributed source, this radius corresponds to $(A/\pi)^{1/2}$
where $A$ is the effective area of the optically thick clouds.
For FSC~15307+3252 at $D_L=4.3$ Gpc and
$S_\nu=510$ mJy in the IRAS 100 $\mu$m band, the black body radius is 
370 pc for $T_{dust}=85$ K.  In Arp~220, 
the 100 $\mu$m black body radius is 180 pc,
which is comparable to the size of the nuclear molecular gas 
complex ($R_{CO}=230$ pc -- \cite{sco97b}).  The size of the CO
emitting region inferred from the gravitational lensing 
model for the Cloverleaf Quasar at $z=2.6$ (H~1413+117; \cite{yun97})
is also similar.  
%Therefore, the observed dust emission 
%in FSC~15307+3252 is likely arising from the nuclear starburst 
%region size scale rather than a circum-nuclear disk scale
%($R\simless 10$ pc).

Both Arp~220 and FSC~10214+4724 are gas rich systems with
molecular gas contents in excess of $10^{10}~M_\odot$ and
gas-to-dust ratios of 260 \& 500, respectively.  These ratios are typical
of luminous infrared galaxies in the IRAS Bright Galaxy Sample 
($M_{H_2}/M_{dust}=540\pm290$; \cite{san91}).  However, given its large 
infrared luminosity it is surprising that no 
CO emission is detected in FSC~15307+3252 by our observations.  
The $3\sigma$ upper 
limit on its CO luminosity implies an upper limit on molecular 
gas mass of $M_{H_2}(3\sigma)\le 8.8\times 10^9 M_\odot$, which
is less than the typical molecular gas content in infrared 
luminous galaxies like Arp~220 even though its FIR luminosity is 
nearly 10 times larger.  The well known relationship between the
FIR and CO luminosity (see review by Young \& Scoville 1991)
has a second order trend of increasing $L_{FIR}/L_{CO}$
[equivalently, $L_{FIR}/M_{H_2}$] ratio with increasing 
infrared luminosity.   The observed large $L_{FIR}/L_{CO}$ ratio
in FSC~15307+3252 is generally consistent with this trend, but 
the observed CO luminosity is still at least 10 times too
small compared to the predicted relation.  Therefore, the
FIR emission in FSC~15307+3252 has a clear and large excess over the
mechanisms operating in other infrared luminous galaxies such
as Arp~220.

A possible explanation for the enhanced FIR luminosity
and apparent large $L_{FIR}/M_{H_2}$ ratio ($>1500$)
is that the mean dust temperature is globally higher in FSC~15307+3252
compared to other infrared luminous galaxies.  The
dust models most consistent with the SED for FSC~15307+3252
have dust temperatures 10 to 20 K warmer than 
in Arp~220 ($T_{dust}=42$ K, $\beta=1.3$
for $\lambda\ge100~\mu$m and $T_{dust}=65$ K for 
$\lambda\le100~\mu$m -- \cite{sco91}).  The flux density
(or luminosity) has an exponential dependence on dust
temperature (see Eq.~3), and a relatively small increase in
mean dust temperature can dramatically increase the observed
FIR luminosity even without any increase in the total dust mass.   
%Increased mean dust temperature 
%in dense neutral gas by X-ray heating in the circum-AGN clouds
%has been suggested by Maloney, Hollenbach, \& Tielens (1996).
The inferred gas-to-dust ratio of $\le$ 
60-250 from the dust model (\S4.1) is somewhat low but still 
consistent with the ratios seen in other galaxies.  

%The inferred FIR-to-CO luminosity ratio is extremely large,
%$L_{60\mu}/L'_{CO}\ge 1900$, although it is typically 100-200 
%for ultraluminous infrared galaxies (Solomon et al. 1997).  

\subsection {Comparison with Infrared Bright Elliptical-like Galaxies}

While Arp~220 and other ultraluminous infrared galaxies in the
local universe are typically ongoing mergers of two gas rich
spiral galaxies (see Hibbard \& Yun 1998), FSC~15307+3252 
appears to be a giant elliptical galaxy, with possible companions 
(\cite{soifer94,evans98}).  While the observed gas and dust
properties of FSC~15307+3252 are consistent with those of infrared
luminous galaxies, the absence of bright CO emission (and 
resulting extreme $L_{FIR}/M_{H_2}$ ratio) and the presence of 
a buried luminous AGN (\cite{hines95,liu96}) 
warrant comparisons with other types of objects.
In particular, we compare the observed properties of FSC~15307+3252
with other infrared bright giant elliptical-like galaxies hosting a
luminous AGN.   

The giant central galaxy of the Perseus cluster, NGC~1275 (Perseus~A,
3C~84), offers an interesting comparison as it is a well known 
infrared source whose cold gas content is also well documented  
($M_{H_2}=6\times10^9~M_\odot$; 
\cite{lazareff89,mirabel89,jaffe90,reuter93,inoue96}).  We have
constructed the SED for this galaxy from the literature and
the OVRO and VLA archival flux databases (the central radio source 
was in its quiescent phase during the 
winter observing season of 1997-1998).  Its 
cold dust content is estimated by subtracting a smooth 
polynomial component (non-thermal contribution from the AGN)
from the observed FIR SED.  As summarized
in Table~3, the observed submillimeter-FIR SED can be modeled
as dust emission arising from relatively cool dust 
($T_{dust}\sim 40$ K) with a total mass of about $2\times10^6~M_\odot$.
The optical luminosity of the underlying galaxy
is nearly identical to that of FSC~15307+3252, but both the
total FIR luminosity and the dust mass are considerably lower
(by factors of 260 and 40, respectively).
High resolution studies of molecular gas and FIR emission have
shown that the associated star forming activity occurs spread 
over a 10 kpc area in NGC~1275 (\cite{reuter93,inoue96,lester95}), 
and this explains the low $L_{FIR}/M_{H_2}$ ratio (8.3) and 
low dust temperature.  Part of the observed infrared emission
may also arise in the foreground system currently 
colliding onto NGC~1275 (see Minkowski 1957, Caulet et al. 1992).

The radio galaxy Cygnus~A offers another interesting comparison 
since it is a giant elliptical galaxy hosting a buried quasar 
nucleus (\cite{djor93,ogle97}).   The optical luminosity of the host
galaxy in Cygnus~A is about 3 times fainter than FSC~15307+3252, but 
most of its total luminosity emerges in the 
infrared ($L_{FIR}/L_B=30$) as in FSC~15307+3252.
The SED of Cygnus~A, excluding the contribution
from the radio lobes, is shown in Figure~3, and the observed
submillimeter-FIR SED can be modeled as thermal emission arising
from about $2\times 10^6~M_\odot$ of dust with $T_{dust}\sim 50$ K.
Again, both the FIR luminosity and total dust mass are
considerably less than those of FSC~15307+3252 (by factors of
76 and 40, respectively).  Previous CO observations have reported 
only upper limits for the molecular gas mass ($\sim10^{10} M_\odot$,
\cite{mirabel89,mazz93,mcnamara94}).  Given its low dust mass, 
at least an order of magnitude improvement in sensitivity
will be needed.  

\subsection {Starburst versus AGN}

In ultraluminous infrared galaxies, the highly 
luminous activity originating
from the central region less than 1 kpc in size is
opaque even at FIR wavelengths ($A_V\ge 10^3$), and a hotly
debated issue is whether the hidden source of luminosity
is a massive starburst or an AGN.  There is little doubt that 
young massive stars are forming within the massive central gas 
complexes in all cases, and evidence for an AGN, such as 
broad emission lines or excess radio emission, is present
some of the luminous infrared galaxies.  
The difficult unresolved issue, however, is whether 
a starburst or an AGN is the dominant source of the observed luminosity.

For the luminous infrared galaxies
where the majority of the bolometric luminosity emerges in the
infrared, the ratio of infrared luminosity to total gas mass
($L_{FIR}/M_{H_2}$) is a measure of efficiency of converting
gas mass into luminosity (\cite{sco91,san91}).  For a starburst
population with a reasonable IMF, the ratio $L/M_{dyn}$ should
be less than $500~L_\odot/M_\odot$ (\cite{leitherer95,sco97b}), and
objects with an $L_{FIR}/M_{H_2}$ ratio $>500~L_\odot/M_\odot$ 
require a significant luminosity contribution from an active nucleus.

Among the galaxies discussed above, only upper limits on the
gas mass exist for FSC~15307+3252 and Cygnus~A.  However,
the ratio $L_{FIR}/M_{dust}$ can still be examined.  
As shown in Table~3,
this ratio for FSC~10214+4724, FSC~15307+3252, and Cygnus~A
is 4 to 15 times larger than for Arp~220 and approaches the limit of
500 $L_\odot/M_\odot$ (assuming a reasonable gas-to-dust ratio 
of $\sim200$), which is difficult to explain with a 
starburst alone.  This result is perhaps not
surprising as all three sources probably host a quasar-like
luminous AGN.  In NGC~1275, the $L_{FIR}/M_{dust}$ ratio
is similar to that of Arp~220, but the situation is not 
clear because the bulk of the gas and dust may lie well
outside the influence of the AGN.

\subsection {Other Dusty High Redshift Galaxies}

Lastly, it is worth noting that there is a growing list of
high redshift galaxies with a clear submillimeter dust continuum 
detection but without correspondingly bright CO emission. This
list includes the z=4.3 radio galaxy 8C~1435+635 (\cite{ivison95,ivison98}),
the z=3.8 radio galaxy 4C~41.17 (\cite{chini94,dunlop94,yun96}), and
the z=2.8 radio galaxy MG~1019+0535 (\cite{yun96,cimatti98}) -- also
see the review by Evans (1997).  Inferred dust masses of about 
$10^8~M_\odot$ are reported for these systems, but their
dust masses are nearly unconstrained because their 
dust continuum detections in most cases are limited to just 
one or two data points near the peak of their dust spectrum 
($\lambda_{rest}\sim 200~\mu$m).  Existing CO observations of 
these systems place rather large gas mass limits of about
$10^{11}~M_\odot$ (\cite{yun96,eva96,vanojik97,ivison98})
-- comparable to the most gas-rich galaxies in the local 
universe but probably not sufficient to detect the gas 
associated with the dust observed at submillimeter
wavelengths.  These observations are still consistent with 
the formation of luminous galaxies in the early epochs.
More sensitive CO observations are needed for
a better comparison of their gas and dust properties.  
\bigskip

\section {Summary and Concluding Remarks}

We report the results of CO (4--3) line and rest frame
650 $\mu$m continuum observations of the z=0.93 hyperluminous
infrared galaxy FSC~15307+3252.  No CO emission was detected
within the redshift range of z=0.9240-0.9275, and we place an
upper limit on the molecular gas mass of $\le 5.0\times10^9
h^{-2} M_\odot$ for this galaxy with $L_{FIR}>10^{13}~L_\odot$.
We have also obtained a $3\sigma$ upper limit on the rest frame 
650 $\mu$m continuum flux of 5.1 mJy.  Combined with 
existing data, an SED which is completely
dominated by the emission in the FIR emerges ($L_{FIR}/L_B=650$).  
Several dust models consistent with the data can be
constructed, and they all imply a
total dust mass associated with the bright FIR emission in
FSC~15307+3252 of 0.4-1.5 times $10^8~M_\odot$.

The gas and dust properties of FSC~15307+3252 are
compared with those of the luminous infrared galaxies
FSC~10214+4724 and Arp~220.
The CO luminosity (molecular gas content)
and gas-to-dust ratio are somewhat low ($M_{gas}/M_{dust}\le$
60-250) but are within the range of observed
properties for more gas-rich infrared galaxies.  
FSC~15307+3252 does not compare well with two infrared bright
elliptical-like galaxies NGC~1275 and Cygnus~A because its  
FIR luminosity and dust content are 100 and 40 times larger,
respectively.

In conclusion, the observed gas and dust properties of the
z=0.93 hyperluminous infrared galaxy FSC~15307+3252 are
consistent with those of a massive galaxy with a
large amount of dust (and possibly gas).  The high dust temperature
and large $L_{FIR}/M_{dust}$ ratio ($\ge10^5$) suggest 
that a large fraction of the luminosity may come from the 
luminous central AGN.   

\acknowledgements

The authors are grateful to A. Evans, T. Phillips, and
M. Gerin for helpful discussions and thank B. Butler for 
carefully reading this manuscript.
The Owens Valley millimeter array is supported by NSF grants 
AST 93-14079 and AST 96-13717.  M. Yun is supported by an
NRAO Jansky Fellowship.  This research has made use of the
NASA/IPAC Extragalactic Database (NED) which is operated by the
Jet Propulsion Laboratory, California Institute of Technology, under
contract with the National Aeronautics and Space Administration.

\clearpage

\figcaption[fig1.ps]{Average spectrum of CO (4--3) emission in
FSC~15307+3252 smoothed to 80 MHz ($\Delta V=100$ \kms) resolution (sampled
at every 40 MHz) is shown for a 3\arcsec\ aperture centered 
on the position of FSC~15307+3252.  The rms noise in each 
80 MHz (100 \kms) channel map is 9 mJy beam$^{-1}$.  The
inferred $3\sigma$ upper limit on total molecular gas mass is
$5.0\times 10^9 h^{-2}~M_\odot$.
\label{fig1}}

\figcaption[fig2.ps]{Spectral Energy Distribution of FSC~15307+3252.
Three ``good" models for dust emission summarized in Table~2 
are shown along with the 200 K black body 
spectrum appropriate for mid- and near-infrared features.
The total dust mass inferred from the maximum (Models A)
and the minimum (Model C) mass models lies between 0.4 and 1.5 times
$10^8~M_\odot$.  The continuum data points for FSC~15307+3252 come from
Cutri et al. (1994), Becker et al. (1995), and this work.  The
filled squares represent detections while the filled triangles represent
$3\sigma$ upper limits.
\label{fig2}}

\figcaption[fig3.ps]{Comparison of spectral energy distributions  
in infrared luminous galaxies and dusty giant elliptical galaxies.  
The data for FSC~10214+4724, NGC~1275, and Cygnus~A are scaled 
for an easier comparison.  The dotted curves
correspond to the dust models described in Table~3.  Photometry
data are assembled from the literature and the NASA/IPAC Extragalactic
Database (NED).  References include: FSC~10214+4724 
(Rowan-Robinson et al. 1991, Telesco 1983); FSC~15307+3252 (this work, 
Cutri et al. 1994, Becker et al. 1995); Arp~220 (Scoville et al. 1991, 
1997b); NGC~1275 (Fich \& Hodge 1991, Lester et al. 1995); 
Cygnus~A (Djorgovski et al. 1993, Robson \& Leeuw 1997).
As in Figure~2, upside-down triangles represent $3\sigma$ upper limits.
\label{fig3}}

\clearpage

\begin{deluxetable}{lll}
%\tablewidth{400pt}
\tablecaption{Summary of CO Observations and the Derived Properties}
\tablehead{
\colhead{}               & 
\colhead{}               & 
\colhead{}      }  
\startdata
RA (B1950) & 15$^h$30$^m$44$s\atop .$6 & \nl
DEC (B1950) & +32$^\circ$52$^\prime$51$^{\prime\prime}_.$0 & \nl
$z_{CO}$ searched & 0.9240-0.9275 & \nl
Luminosity Distance\tablenotemark{a} &  4.33 Gpc & \nl
Angular-Size Distance\tablenotemark{a} & 1.16 Gpc \hskip 0.3in (1$^{\prime\prime} \rightarrow$ 5.6 kpc) & \nl
$\theta$ (FWHM) & $2''.8 \times 3''.8$ & \nl
$S_{102GHz} (3\sigma)$ & $\le1.3$ mJy & \nl
$S_{239GHz} (3\sigma)$ & $\le5.1$ mJy & \nl
$S_{CO}\Delta V (3\sigma)$\tablenotemark{b} & $\le1.6$ Jy \kms & \nl
$L'_{CO} (3\sigma)$\tablenotemark{b} & $\le2.2\times10^9$ K \kms pc$^2$ & \nl
$M_{H_2} (3\sigma)$\tablenotemark{c} & $\le8.8\times10^{9}$ M$_\odot$ & \nl
%$L_{60\mu}$ & $4.0\times10^{12}$ L$_\odot$ & \nl
%$L_{60\mu}/M_{H_2}$ & $\ge450~L_\odot /M_\odot$ & \nl
\tablenotetext{a}{H$_o$ = 75 km s$^{-1}$ Mpc$^{-1}$, q$_o$ = 0.5}
\tablenotetext{b}{assuming $\Delta V = 300$ \kms}
\tablenotetext{c}{using a standard conversion $\alpha$=4 M$_\odot$ 
(K km s$^{-1}$ pc$^2$)$^{-1}$}
\enddata
\end{deluxetable}

%\clearpage

\begin{deluxetable}{cccccc}
\tablecaption{Dust Models for FSC~15307+3252}
\tablehead{ &
\colhead{Model }               & \colhead{$T_{dust}$}  &
\colhead{$\beta$\tablenotemark{\dagger}}   & \colhead{$M_{dust}$} & }  
\startdata
& Model A & 35 K & 2.0 & $1.5\times10^8~M_\odot$ & \nl
& Model B & 50 K & 1.5 & $9.2\times10^7~M_\odot$ & \nl
& Model C & 65 K & 1.5 & $3.5\times10^7~M_\odot$ & \nl
\tablenotetext{\dagger}{$\beta$ is emissivity index, i.e. $Q(\nu) \propto \nu^\beta$ (see Eq.~1).}
\enddata
\end{deluxetable}

\clearpage

\begin{deluxetable}{lccccc}
\tablecaption{Comparison of Infrared Bright Galaxies}
\tablehead{
\colhead{}               & \colhead{F10214+4724$^a$}     &
\colhead{F15307+3252}    & \colhead{Arp 220}             & 
\colhead{NGC 1275}       & \colhead{Cygnus A}      }  
\startdata
$D_L^b$ (Mpc) & 11800 & 4330 & 77  & 72 &  224 \nl
log $L_B$ ($L_\odot$)$^b$ & 12.10 & 10.31 & 10.05 & 10.53 &  9.78 \nl
log $L_{FIR}$ ($L_\odot$)$^c$ & 14.27 & 13.12 & 12.12 & 10.70 & 11.24 \nl
log $M_{H_2}$ ($M_\odot$)$^d$ & 11.34 & $<$9.95 & 10.28 & 9.78 & $<$9.54 \nl
$T_{dust}$ (K)$^e$, $\beta$ & 75, 1.3 & 50, 1.5 & 42, 1.3 & 40, 1.5 & 50, 1.5 \nl
log $M_{dust}$ ($M_\odot$)$^f$ & 8.64 & 7.54-8.18 & 7.87 & 6.25 & 6.35 \nl
$L_{FIR}/L_B$  & 150 & 650 & 120 & 1.5 & 30 \nl
$L_{FIR}/M_{H_2}$  & 850 & $>$1500 & 70 & 8.3 & $>$50 \nl
$M_{H_2}/M_{dust}$ & 500 & $<$60-250 & 260 & 3400 & $<$1500 \nl
$L_{FIR}/M_{dust}$ & $4.3\times10^5$ & 0.9-3.8$\times10^5$ & $1.8\times10^4$ & $2.8\times10^4$ & $7.8\times10^4$ \nl
\tablenotetext{a}{magnification due to gravitational lensing not included.}
\tablenotetext{b}{log $L_B = 11.898 - 0.4m_B + 2~{\rm log}~D_L$, where
$D_L$ is the luminosity distance in Mpc assuming
H$_o$ = 75 km s$^{-1}$ Mpc$^{-1}$, q$_o$ = 0.5}
\tablenotetext{c}{computed from IRAS 60 \& 100 $\mu$m fluxes for Arp~220,
NGC~1275, \& Cygnus~A (see Helou et al. 1988) and computed from the
adopted dust model SEDs for FSC~10214+4724 and FSC~15307+3252.}
\tablenotetext{d}{using $\alpha$=4 M$_\odot$ (K km s$^{-1}$ pc$^2$)$^{-1}$}
\tablenotetext{e}{derived from the dust models shown in Figure 3.}
\tablenotetext{f}{see Eq.~3.}
\enddata
\end{deluxetable}

\clearpage
\plotone{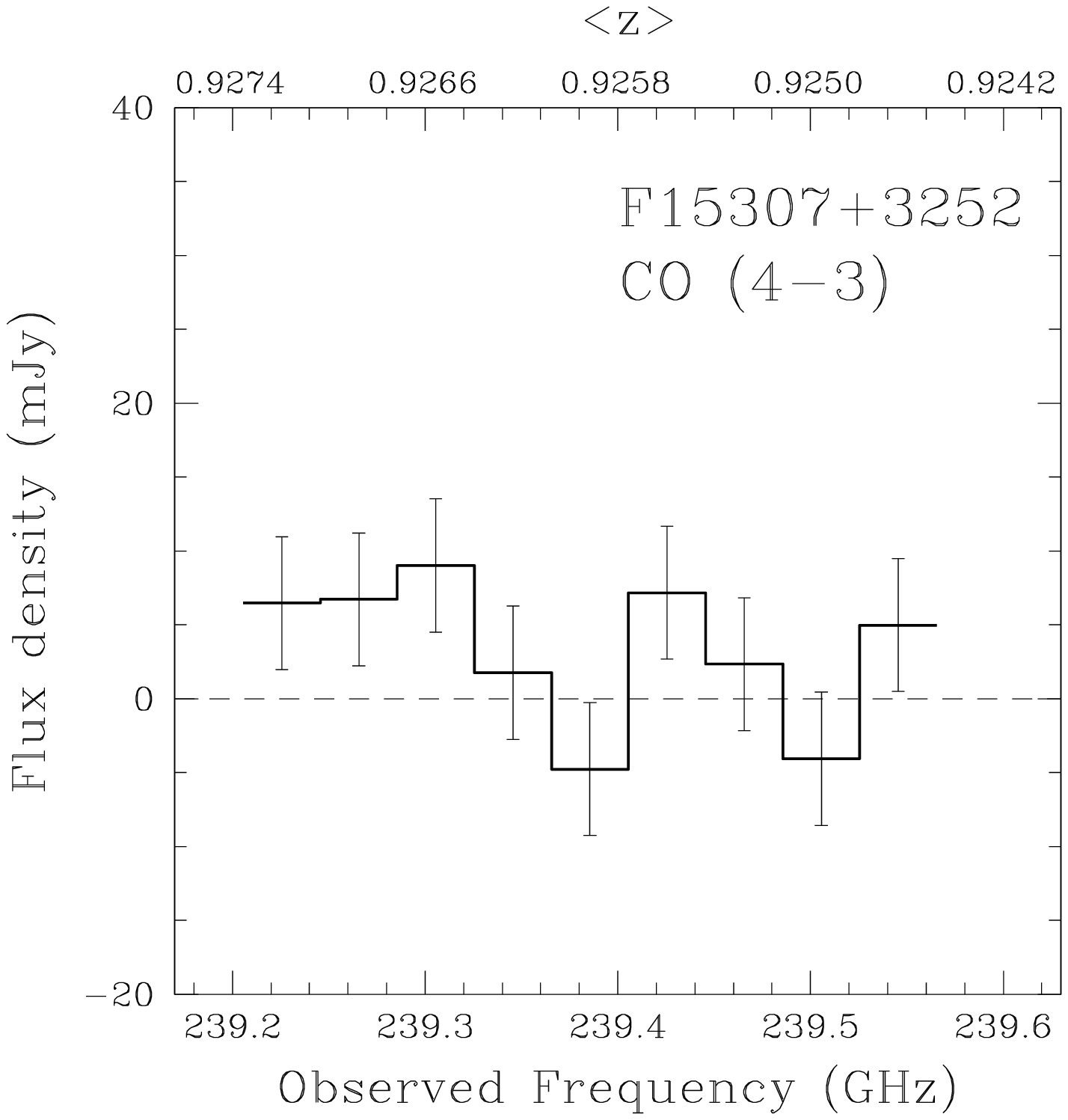}

\clearpage
\plotone{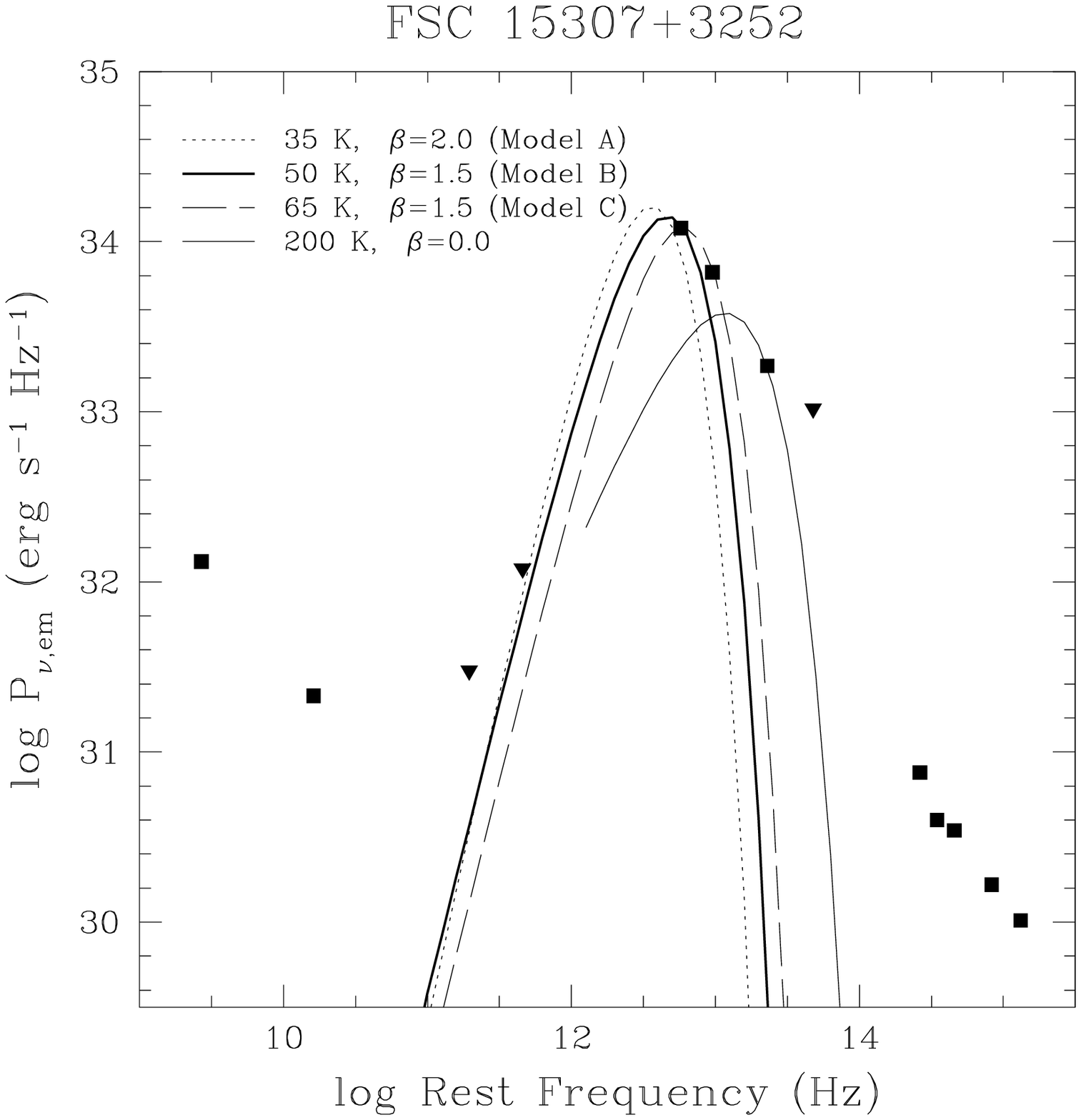}

\clearpage
\plotone{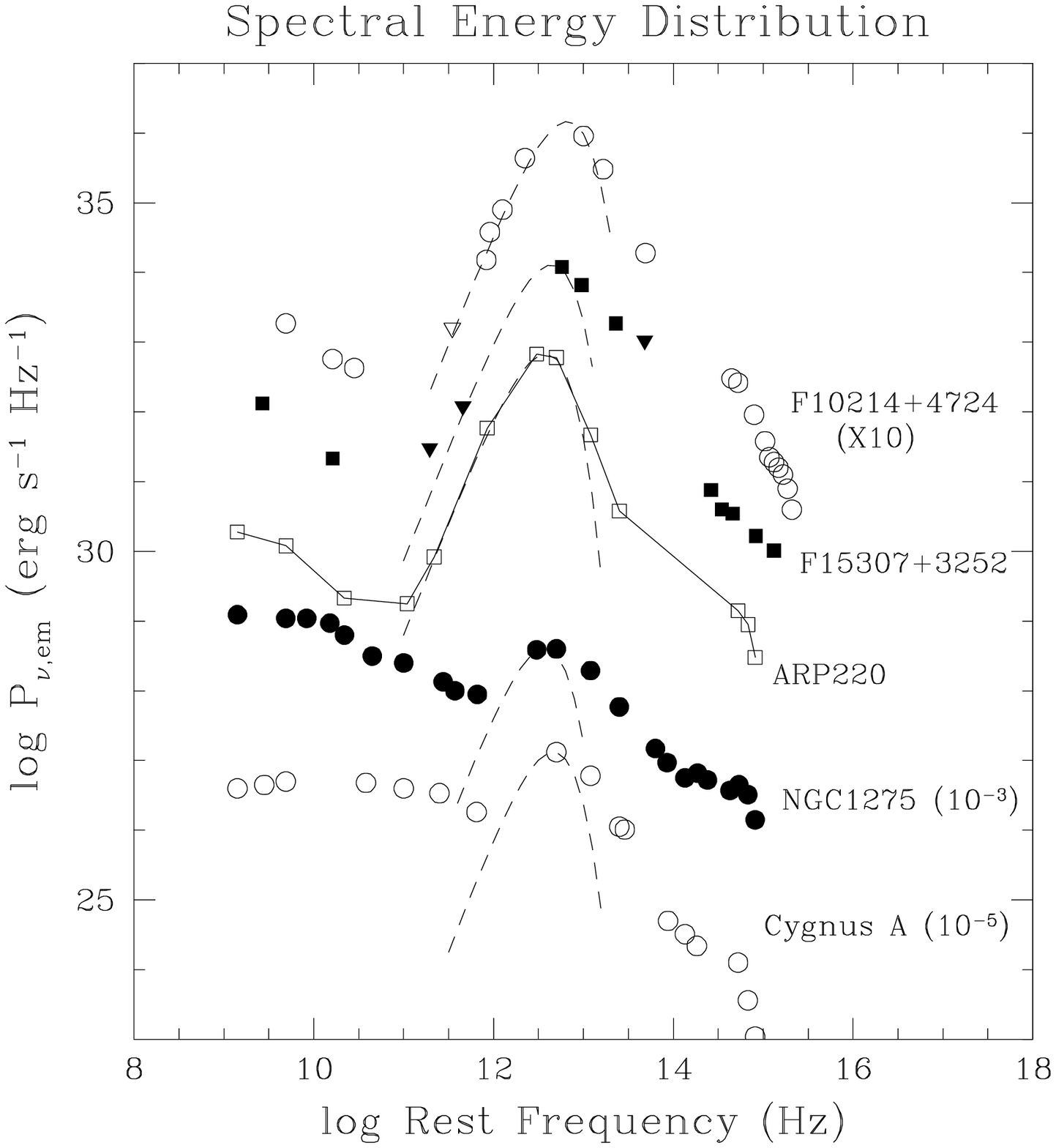}


\clearpage

\begin{thebibliography}{}

\bibitem[Becker et al. 1995]{becker95} Becker, R.H., White, R.L., 
	\& Helfand, D.J. 1995, ApJ, 450, 559

%\bibitem[Becker et al. 1997]{becker97} Becker, R.H., Gregg, M.D., Hook, 
%	I.M., McMahon, R.G., White, R.L., Helfand, D.J. 1997, ApJ, 479, L93

\bibitem[Brown \& Vanden Bout 1991]{bv91} Brown, R., \& Vanden Bout, P.
	1991, \aj, 102, 1956

%\bibitem[Brown \& Vanden Bout 1992]{bv92} Brown, R., \& VandenBout, P. 
%	1992, \apj, 397, L19 (extended CO in I10214)

%\bibitem[Brown \& Vanden Bout 1993]{bv93} Brown, R., \& VandenBout, P. 
%	1993, \apj, 412, L21 

\bibitem[Caulet et al. 1992]{cau92} Caulet, A., Woodgate, B. E., Brown,
	L. W., Gull, T. R., Hintzen, P. et al. 1992, ApJ, 388, 301

%\bibitem[Chambers et al. 1990]{chambers90} Chambers, K. C., Miley, G. K., 
%	\& van Breugel, W. J. M. 1990, \apj, 363, 21

\bibitem[Chini \& Kr\"{u}gel 1994]{chini94} Chini, R., \& Kr\"{u}gel, E.
	1994, A\&A, 288, L33

\bibitem[Cimatti et al. 1998]{cimatti98} Cimatti, A., Freudling, W.,
	R\"{o}ttgering, H. J. A., Ivison, R. J., \& Mazzei, P. 1998,
	A\&A, 329, 399

\bibitem[Cutri et al. 1994]{cutri94} Cutri, R.M., Huchra, J.P., Low, 
	F.J., Brown, R.L., Vanden Bout, P.A. 1994, ApJ, 424, 65

\bibitem[Djorgovski et al. 1991]{djor91} Djorgovski, S., Weir, N., 
	Matthews, K., \& Graham, J. R. 1991, ApJ, 372, L67 

\bibitem[Djorgovski et al. 1993]{djor93} Djorgovski, S., \& Thompson, D. 1993, in {\it IAU Symp. 149,
The Stellar Populations in Galaxies}, eds. Renzini, A., Barbuy, B.
(Kluwer, Dordrecht), p337

\bibitem[Downes et al. 1997]{downes97} Downes et al. 1997, in {\it Highly 
	Redshifted Radio Lines}, eds. C. Carilli, S. Radford, K.
	Menten, \& G. Langston, in press.


%\bibitem[Downes et al. 1995]{dow95} Downes, D., Solomon, P. M. \& 
%	Radford, S. E. 1995, \apjl, 453, L65.

%\bibitem[Draine \& Lee 1984]{draine84} Draine, B.T., \& Lee, H.M. 1984,
%	ApJ, 285, 89

\bibitem[Dunlop et al. 1994]{dunlop94} Dunlop, J. S., Hughes, D. H.,
	Rawlings, S., Eales, S. A., \& Ward, M. J. 1994, Nature, 370, 347

%\bibitem[Eales \& Rawlings 1993]{er93} Eales, S. A., \& Rawlings, S. 
%1993, \apj, 411, 67 

\bibitem[Evans 1997]{evans97} Evans, A. S., in {\it Highly 
	Redshifted Radio Lines}, eds. C. Carilli, S. Radford, K.
	Menten, \& G. Langston, in press.

\bibitem[Evans et al. 1996]{eva96} Evans, A. S., Sanders, D. B., 
	Mazzarella, J. M., Solomon, P. M., Downes, D., Kramer, C.,
	\& Radford, S. J. E., 1996, \apj, 457, 658

\bibitem[Evans et al. 1998]{evans98} Evans, A. S., Sanders, D. B.,
	Cutri, R. M., Radford, S. J. E., Surace, J. A. et al. 1998, \apj, in press.

\bibitem[Fich \& Hodge 1991]{fich91} Fich, M., \& Hodge, P. 1991,
	ApJ, 374, L17

\bibitem[Fischer et al. 1998]{fischer98} Fischer et al. 1998, in
	{\it Extragalactic Astronomy in the Infrared}, eds. G.A.
	Mamon, T.X. Thuan, \& J.T.T. Van, (Editions Frontieres,
	Gif-sur-Yvette), in press.

\bibitem[Hacking et al. 1997]{hacking97} Hacking, P., Gautier, T. N.,
	Herter, T. L., Lonsdale, C. J., Shupe, D. L. et al. 1997, in
	{\it Extragalactic Astronomy in the Infrared,} eds. G. A. Mamon,
	T. X. Thuan, J. Tran Thanh Van (Editions Frontieres)

\bibitem[Helou et al. 1988]{hel88} Helou, G., Khan, I. R., Malek, L.,
\& Boehmer, L. 1988, ApJS, 68, 151

\bibitem[Hibbard \& Yun 1998]{hib98} Hibbard, J. E., \& Yun, M. S. 1998, in preparation

\bibitem[Hildebrand 1983]{hilde83} Hildebrand, R.H. 1983, QJRAS, 24, 267

\bibitem[Hines et al. 1995]{hines95} Hines, D.C., Schmidt, G.D., Smith, 
	P.S., Cutri, R.M., Low, F.J. 1995, ApJ, 450, 1

\bibitem[Hughes et al. 1993]{hughes93} Hughes, D. H., Robson, E. I.,
	Dunlop, J. S., \& Gear, W. K. 1993, MNRAS, 263, 607

\bibitem[Inoue et al. 1996]{inoue96} Inoue, M. Y., Kameno, S., Kawabe, R.,
	Inoue, M., Hasegawa, T., \& Tanaka, M. 1996, AJ, 111, 1852

\bibitem[Ivison 1995]{ivison95} Ivison, R. J. 1995, MNRAS, 275, L33

\bibitem[Ivison et al. 1998]{ivison98} Ivison, R. J., Dunlop, J. S., 
	Hughes, D. H., Archibald, E. N., Stevens, J. A., Holland, W. S.,
	Robson, E. I., Eales, S. A., Rawlings, S., Dey, A., \& Gear, W. K.
	1998, ApJ, 494, 211

\bibitem[Jaffe 1990]{jaffe90} Jaffe, W. 1990, A\&A, 240, 254

%\bibitem[Keel \& Windhorst 1993]{kw93} Keel, W. C., \& Windhorst, R. A. 1993, %\aj, 106, 455 

\bibitem[Kr\"{u}gel et al. 1990]{krugel90} Kr\"{u}gel, E., Steppe, H.,
	\& Chini, R. 1990, A\&A, 299, 17

\bibitem[Lazareff et al. 1989]{lazareff89} Lazareff, B., Castets, A.,
	Kim, D.-W., \& Jura, M. 1989, ApJ, 336, L13 

\bibitem[Leitherer \& Heckman 1995]{leitherer95} Leitherer, C., \&
	Heckman, T. M. 1995, ApJS, 96, 9

\bibitem[Lester et al. 1995]{lester95} Lester, D. F., Zink, E. C.,
	Doppmann, G. W., Gaffney, N. I., Harvey, P. M., Smith, B. J.,
	\& Malkan, M. 1995, ApJ, 439, 185 

\bibitem[Liu et al. 1996]{liu96} Liu, M.C., Graham, J.R., Wright, G.S. 
	1996, ApJ, 470, 771

\bibitem[Malhotra et al. 1997]{malhotra97} Malhotra et al. 1997, ApJ, 419, L27
	
%\bibitem[Maloney et al. 1996]{malony96} Maloney, P.R., Hollenbach, D.J.,
%	\& Tielens, A.G.G.M. 1996, ApJ, 466, 561

\bibitem[Mazzarella et al. 1993]{mazz93} Mazzarella, J.M., Graham, J.R.,
	Sanders, D.B., \& Djorgovski, S. 1993, ApJ, 409, 170 
	
%\bibitem[McCarthy et al. 1991]{mc91} McCarthy, P. J., van Breugel, W., 
%Kapahi, V., \& Subrahmanya, C. 1991, \aj, 102, 522

\bibitem[McNamara \& Jaffe 1994]{mcnamara94} McNamara, B.R., \&
	Jaffe, W. 1994, A\&A, 281, 673 

\bibitem[Minkowski 1957]{min57} Minkowski, in IAU Symp. 4, {\it
	Radio Astronomy}, ed. H. C. van de Hulst (Cambridge: Cambridge
	University Press), p107

%\bibitem[Mirabel \& Sanders 1989]{ms89} Mirabel, I. F. and Sanders, 
%	D. B. 1989, \apj, 340, 253 

\bibitem[Mirabel et al. 1989]{mirabel89} Mirabel, I. F., Sanders, D. B.,
	\& Kazes, I. 1989, ApJ, 340, L9 

%\bibitem[Neuschaefer \& Windhorst 1995]{nw95} Neuschaefer, L. W., 
%	\& Windhorst, R. A. 1995, \apjs, 96, 371 

\bibitem[Ogle et al. 1997]{ogle97} Ogle, P. M., Cohen, M. H., Miller, J. S.,
	Tran, H. D., Fosbury, R. A. E., \& Goodrich, R. W. 1997, ApJ, L37

\bibitem[Ohta et al. 1996]{ohta96} Ohta, K., Yamada, T., Nakanishi, K., 
	Kohno, K., Akiyama, M. \& Kawabe, R. 1996, Nature, 382, 426

\bibitem[Omont et al. 1996]{omont96} Omont, A., Petitjean, P., Guilloteau, 
	S., McMahon, R. G. \& Solomon, P. M. 1996, Nature, 382, 428

\bibitem[Reuter et al. 1993]{reuter93} Reuter, H.-P., Pohl, M., Lesch, H.,
	\& Sievers, A.W. 1993, A\&A, 277, 21 

\bibitem[Robson \& Leeuw 1997]{robson97} Robson, E. I., \& Leeuw, L. 1997,
	BAAS, 191, 22.01

\bibitem[Rowan-Robinson et al. 1991]{rowan91} Rowan-Robinson, M.
	Broadhurst, T., Lawrence, A., McMahon, R.G., Lonsdale, C.J.,
	Oliver, S.J., Taylor, A.N., Hacking, P.B., Conrow, T., Saunders, W.,
	Ellis, R.S., Efstathiou, G.P., Condon, J.J. 1991, Nature, 351, 719

\bibitem[Sanders et al. 1991]{san91} Sanders, D. B., Scoville, N.Z., 
	\& Soifer, B. T. 1991, \apj, 370, 158

\bibitem[Scoville et al. 1991]{sco91} Scoville, N. Z., Sargent, A. I,
	Sanders, D. B., \& Soifer, B. T. 1991, ApJ, 366, L5

\bibitem[Scoville et al. 1992]{sco92} Scoville, N. Z., Carlstrom, 
	J. C., Chandler, C. J., Phillips, J. A., Scott, S. L., Tilanus, 
	R. P., \& Wang, Z. 1992, PASP, 105, 1482

%\bibitem[Scoville et al. 1994]{sco94} Scoville, N. Z., Hibbard, J. E,
%	Yun, M. S., \& van Gorkom, J. H. 1994, in {\it Mass Transfer
%	Induced Activities in Galaxies,} ed. I Shlosman (Cambridge
%	University Press: Cambridge)

\bibitem[Scoville et al. 1995]{sco95} Scoville, N. Z., Yun, M. S., Brown, 
	R. L. \& VandenBout, P. A. 1995, \apj, 449, L109

\bibitem[Scoville et al. 1997a]{sco97a} Scoville, N.Z., Yun, M.S., 
Windhorst, R.A., Keel, W.C., Armus, L. 1997a, ApJ, 485, L21

\bibitem[Scoville et al. 1997b]{sco97b} Scoville, N. Z., Yun, M. S. 
	\& Bryant, P. 1997b, \apj, 484, 702

\bibitem[Shepherd et al. 1994]{she94} Shepherd, M. C., Pearson, T. J., 
	\& Taylor, G. B. 1994, BAAS, 26, 987

\bibitem[Smail et al. 1997]{smail97} Smail, I. Ivison, R. J., \& Blain,
	A. W. 1997, \apjl, 490, L5

\bibitem[Soifer et al. 1994]{soifer94} Soifer, B. T., Neugebauer, G.,
	Matthews, K., \& Armus, L. 1994, ApJ, 433, L69

\bibitem[Solomon et al. 1992a]{sol92a} Solomon, P. M., Downes, D., 
	\& Radford, S. J. E. 1992a, Nature,  356, 318

\bibitem[Solomon et al. 1992b]{sol92b} Solomon, P. M., Downes, D., \& 
	Radford, S. J. E. 1992b, \apj, 398, L29

\bibitem[Solomon et al. 1997]{sol97} Solomon, P.M., Downes, D., Radford, 
S.J.E., Barrett, J.W. 1997, ApJ, 478, 144

\bibitem[Telesco 1993]{telesco93} Telesco, C.M. 1993, MNRAS, 263, L37
	
\bibitem[van Ojik et al. 1997]{vanojik97} van Ojik, R., R\"{o}ttgering,
	H. J. A., van der Werf, P. P., Miley, G. K., Carilli, C. L.,
	Visser, A., Isaak, K. G., Lacy, M., Jenness, T., Sleath, J.,
	\& Wink, J. 1997, A\&A, 321, 389

\bibitem[Young \& Scoville 1991]{you91} Young, J. S. \& Scoville,
	N. Z. 1991, ARAA, 29, 581

\bibitem[Yun \& Scoville 1996]{yun96} Yun, M. S., \& Scoville, N. Z.
	1996, in {\it IAU Symposium 170, CO: 25 Years of Millimeter
	Spectroscopy}, eds. W. B. Latter, S. J. E. Radford, P. R.
	Jewell, J. G. Mangum, \& J. Bally (Kluwer: Dordrecht), p341

\bibitem[Yun et al. 1997]{yun97} Yun, M. S., Scoville, N. Z., 
	Carrasco, J. J., \& Blandford, R. D. 1997, \apj, 479, L9 


\end{thebibliography}
\end{document}